\documentclass{jps-cp}
\usepackage{txfonts}
\usepackage{graphicx,amsmath,amssymb,wrapfig}
\title{Conference Summary of QNP2018}
\author{Shunzo \textsc{Kumano}}
\inst{
KEK Theory Center, Institute of Particle and Nuclear Studies, KEK, \\
\ \ Oho 1-1, Tsukuba, Ibaraki, 305-0801, Japan \\
J-PARC Branch, KEK Theory Center,
Institute of Particle and Nuclear Studies, KEK \\
\ \ and Theory Group, Particle and Nuclear Physics Division, 
J-PARC Center, \\
\ \ Shirakata 203-1, Tokai, Ibaraki, 319-1106, Japan \\
}

\email{shunzo.kumano@kek.jp}

\recdate{February 19, 2019}

\abst{This report is the summary of the Eighth International Conference 
on Quarks and Nuclear Physics (QNP2018). 
Hadron and nuclear physics is the field to investigate 
high-density quantum many-body systems bound 
by strong interactions. It is intended to clarify 
matter generation of universe and properties of 
quark-hadron many-body systems.
The QNP is an international conference which covers a wide range
of hadron and nuclear physics, including quark and gluon structure 
of hadrons, hadron spectroscopy, hadron interactions 
and nuclear structure, hot and cold dense matter, 
and experimental facilities.
First, I introduce the current status of the hadron 
and nuclear physics field related to this conference. 
Next, the organization of the conference is explained, 
and a brief overview of major recent developments is discussed
by selecting topics from discussions at the plenary sessions.
They include rapidly-developing field of
gravitational waves and nuclear physics,
hadron interactions and nuclear structure with strangeness,
lattice QCD, hadron spectroscopy, nucleon structure, heavy-ion physics,
hadrons in nuclear medium, and experimental facilities of 
EIC, GSI-FAIR, JLab, J-PARC, Super-KEKB, and others.
Nuclear physics is at a fortunate time to push various projects
at these facilities. However, we should note that
the projects need to be developed together with related studies
in other fields such as gravitational physics,
astrophysics, condensed-matter physics, particle physics,
and fundamental quantum physics.}

\kword{QNP2018 conference, Quarks, Hadrons, Nuclei, QCD}

\begin{document}
\maketitle

\section{Introduction to nuclear physics at QNP2018}

\begin{wrapfigure}[11]{r}{0.38\textwidth}
   \vspace{-0.9cm}
   \hspace{+1.00cm}
   \begin{center}
   \hspace{-0.50cm}
     \includegraphics[width=5.8cm]{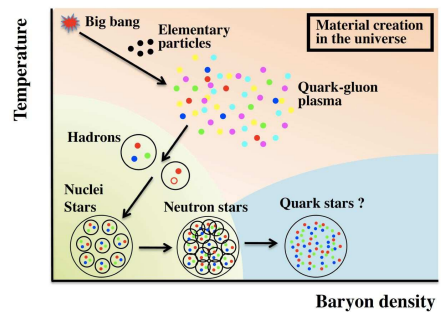}
   \end{center}
\vspace{-0.35cm}
\caption{\hspace{-0.25cm} QCD phase diagram.}
\label{fig:qcd-phase}
\vspace{-0.7cm}
\end{wrapfigure}

Hadron and nuclear physics is the field to investigate properties
of many-body systems bound by strong interactions and 
to understand matter generation of universe starting from
big bang, then developments to quark and gluon plasma, hadrons,
nuclei, and further to neutron stars 
as illustrated in Fig.\,\ref{fig:qcd-phase} \cite{intro-sk}. 
The wide range of topics on the 
quantum-chromodynamics (QCD) phase diagram, such as
quark-gluon plasma, dense stars, and hadrons in nuclear medium,
are discussed in the QNP2018 session, ``hot and cold dense matter".
Hereafter, the terminology ``nuclear physics" is often used in this article
by taking it as a broad field including hadron physics.
Nuclear physics started as a field to study nuclear structure and reactions.
The basic constituents of nuclei are protons, neutrons, and mesons
which mediate the nuclear force. Nucleon-nucleon (NN) interactions have been
determined by NN scattering measurements and deuteron properties.
Nuclear structure and reactions have been investigated by using
these basic NN interactions. Now, new types of nuclei 
are investigated by extending flavor degrees freedom to strange and charm.
Especially, hypernuclear experiments are now in progress.
We know that peculiar properties of nuclei, as a dense many-body system,
often stem from the Pauli exclusion principle.
For example, independent particle models work as good descriptions
of nuclei in spite of the fact that the average nucleon separation 
is almost equal to the diameter of the nucleon.
We are now stepping into nuclei partially without the exclusion principle 
by hadrons with strangeness and charm.
Furthermore, there are significant progress recently
from the lattice QCD side on baryon interactions.
These topics were discussed in the session,
``hadron interactions and nuclear structure".

Hadron properties have been investigated by measuring global observables
such as masses, decay widths, spins, and parities. In comparison with
theoretical calculations on these quantities, several exotic-hadron 
candidates have been reported experimentally in the last decade.
It is a fortunate time to investigate hadron-structure physics 
by extending our knowledge to exotic states beyond the original
quark model of 1964 \cite{quark-model}. 
The topics of exotic hadrons and N$^*$ resonances
are discussed in the session, ``hadron spectroscopy".
On the other hand, parton structure of nucleon
and nuclei has been studied in deep inelastic scattering of leptons
and also hadron-hadron collisions. It is now coming to the stage
of probing three-dimensional (3D) structure functions which contain
information on form factors and parton distribution functions.
The 3D tomography is crucial to clarify the origin of nucleon
spin including partonic orbital-angular-momentum contributions
and to open a new field of gravitational form factors of hadrons.
These topics were discussed in the session, ``quark and 
gluon structure of hadrons".

\begin{figure}[b!]
\vspace{-0.50cm}
\begin{minipage}{\textwidth}
\begin{tabular}{lc}
\hspace{-0.30cm}
\begin{minipage}[c]{0.375\textwidth}
   \vspace{-0.2cm}
   \begin{center}
    \includegraphics[width=5.0cm]{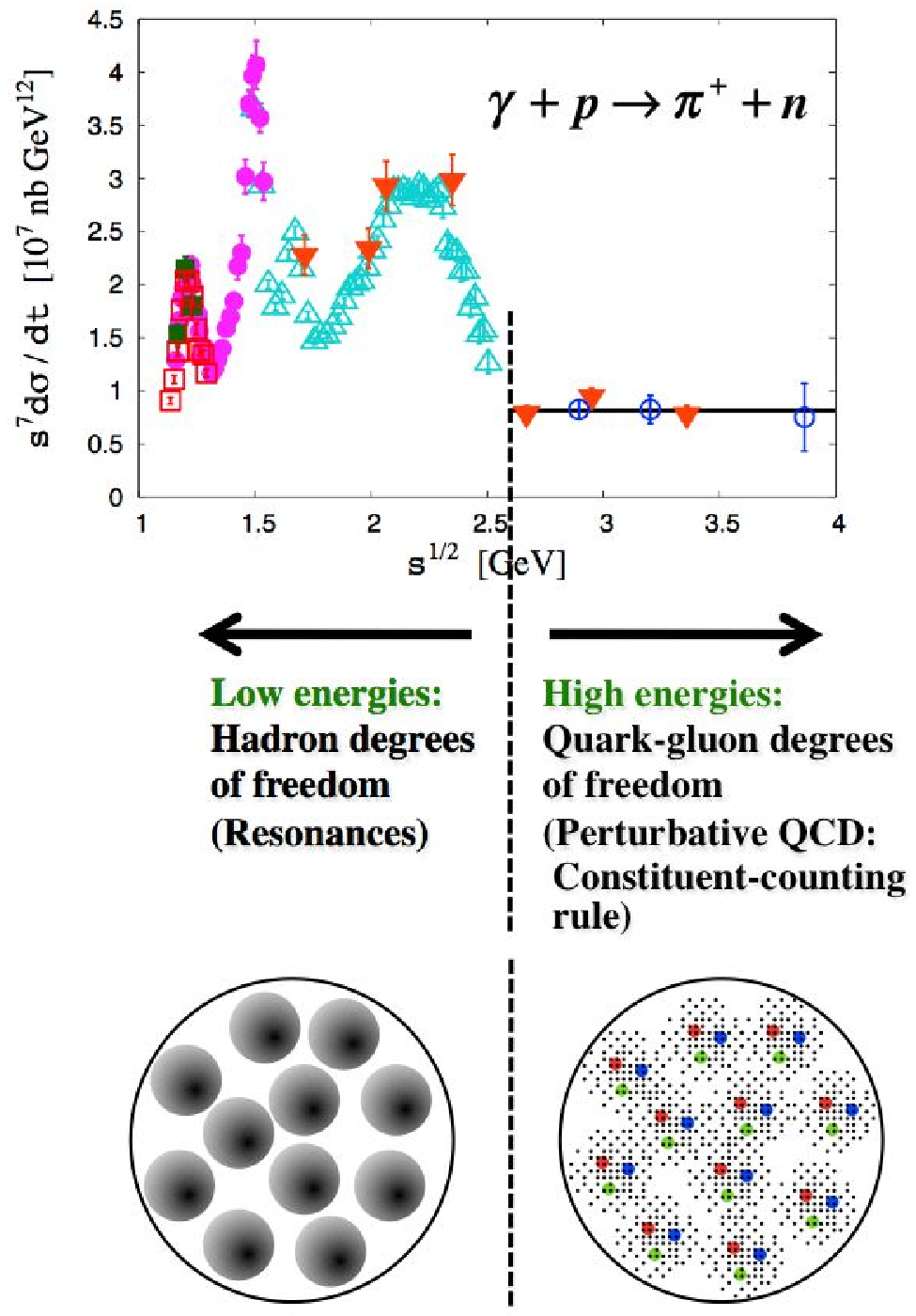}
   \end{center}
\vspace{-0.20cm}
\caption{\hspace{-0.30cm}
Descriptions in terms of hadron 
         or quark-gluon degrees of freedom.}
\label{fig:quak-hadron-transition}
\vspace{-0.4cm}
\end{minipage} 
\hspace{0.3cm}
\begin{minipage}[c]{0.60\textwidth}
    \vspace{-0.3cm}
    \hspace{1.20cm}
    \includegraphics[width=6.8cm]{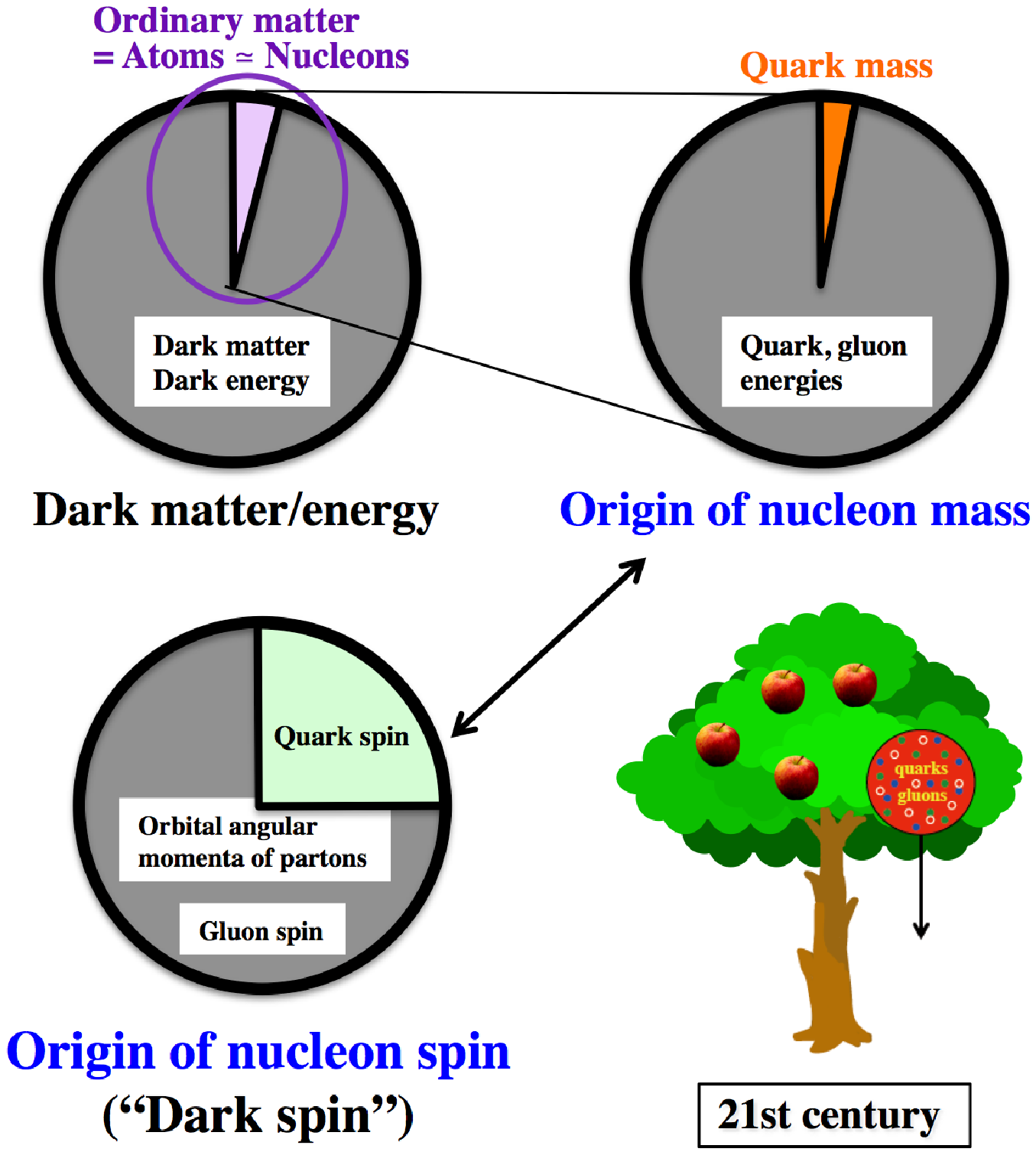}
\vspace{-0.00cm}
\caption{\hspace{-0.30cm}
Origin of hadron masses in terms of quarks and gluons.}
\label{fig:hadron-masses}
\vspace{-0.4cm}   
\end{minipage}
\end{tabular}
\vspace{0.20cm}
\end{minipage}
\end{figure}

Hadrons, nuclei, quark-gluon plasma, and neutron stars are
bound systems of strong interactions, and the basic theory
of strong interactions is known as QCD. 
All these matters should be described by the framework
of QCD, namely by quark and gluon degrees of freedom, in principle.
However, it is not straightforward to apply QCD to these topics
in the whole kinematical region.
At high energy reactions with large-momentum transfer, we know 
that the perturbative QCD can be applied because of 
the asymptotic freedom. At low energies, 
we rely on lattice QCD simulations; however, it is still very limited
and it cannot be used at finite baryon densities. 
Therefore, it is useful and often the only method
to employ descriptions of effective hadron and nuclear models
in terms of hadron degrees of freedom \cite{intro-sk}.
An example is shown in Fig.\,\ref{fig:quak-hadron-transition} \cite{counting}
to indicate a possible transition from
hadron degrees of freedom to quark-gluon degrees of freedom
by the exclusive reaction $\gamma + p \to \pi^+ + n$
with the scattering angle of 90$^\circ$
in the center-of-mass frame.
In hard exclusive hadron reactions, it is known that
cross sections follow the constituent counting rule,
namely the straight line in Fig.\,\ref{fig:quak-hadron-transition}.
Although the transition energy could change depending
on reactions, nuclei are described by hadron degrees of freedom
at low energies and by quark-gluon ones at high energies.

One of major fundamental nuclear-physics topics is to understand
how hadron masses are generated because 
up- and down-quark masses are very small 
as illustrated in Fig.\,\ref{fig:hadron-masses} \cite{hadron-mass-1,sf-summary}.
It is a similar topic to investigate the origin of nucleon spin.
The nucleon spin is given by the matrix element of 
total angular momentum expressed by the quark-gluon energy-momentum
tensor, whereas the hadron mass is expressed by
the quark-gluon energy-momentum tensor.
The scalar component of the matrix element of the energy-momentum tensor
is the source of the gravity, namely the hadron mass, and 
the vector component indicates pressure and shear-forces
inside the hadron  \cite{hadron-mass-1,sf-summary}.
This hadron tomography is a fast developing field, 
and time has come to understand ``how an apple falls in terms
of quarks and gluons".

\section{QNP2018 organization}
\vspace{-0.10cm}

The QNP conferences are series of meetings previously held in Adelaide (2000), 
J\"ulich (2002), Bloomington (2004), Madrid (2006), Beijing (2009), 
Palaiseau (2012), and Valparaiso (2015). 
This QNP2018 was held at the Tsukuba International Congress Center
in Tsukuba, Japan on November 13-17, 2018.
At this conference, experimentalists and theorists discussed recent 
developments in nuclear physics, 
and the following topics were covered \cite{qnp2018}: \\ \hspace{1.4cm} 
(A) Quark and gluon structure of hadrons, \\ \hspace{1.4cm} 
(B) Hadron spectroscopy, \\ \hspace{1.4cm}  
(C) Hadron interactions and nuclear structure, \\ \hspace{1.4cm}  
(D) Hot and cold dense matter. \\
There were also discussions on future facilities and presentations
on fundamental physics experiments by using the neutron and nuclei.
In addition to plenary sessions, parallel sessions were organized 
in these topics. A poster session was arranged. 
There are 24 talks in the plenary session, 128 (A:27, B:36, C:27, D:38)
talks in the parallel, and 40 (A:6, B:9, C:7, D:16, other:2) posters.

We had 216 participants (students: 45, postdocs: 44, staffs: 127).
There are one participant from Africa, 162 from Asia, 
32 from Europe, 16 from North America, and 5 
from South America: \\ \hspace{1.4cm}    
Africa 1 (South Africa:\,1), \\ \hspace{1.4cm}  
Asia 162 (China:\,18, India:\,3, Indonesia:\,4, Israeli:\,4,
          Japan:\,114,\,Korea:\,12, Taiwan:\,6,  \\ \hspace{3.0cm} 
          Turkey:\,1), \\ \hspace{1.4cm}   
Europa 32 (Austria:\,3, Belgium:\,1, France:\,4, Germany:\,11, Hungary:\,1,
           Italy:\,1, Poland:\,1,  \\ \hspace{3.25cm}  
           Portugal:\,1, Russia:\,5, Slovenia:\,1, Spain:\,2, UK:\,1),
            \\  \hspace{1.4cm} 
North America 16 (Mexico:\,2, USA:\,14),  \\ \hspace{1.4cm} 
South America 5 (Brazil:\,2, Chile:\,3).

The members of the international advisory committee are \\
S. Brodsky, W. Brooks, V. Burkert, W.-C. Chang, H. Enyo, 
A. Gal, H. Gao, M. Garcon, P. Giubellino, T. Hatsuda,
B. Kopeliovich, T.-S.~H.~Lee, S. H. Lee, M. Lutz, Y.-G. Ma, 
G. Martinez, R. McKeown, L. McLerran, C. Meyer, A.~K.~Mohanty, 
T. Nagae, S. Nagamiya, T. Nakano, M. Oka, E. Oset, 
B.~Pasquini, J.-C. Peng, B. Pire, J. Qiu, B. Sharkov,
I. Strakovsky, M. Strikman, H. Tamura, U. Thoma, A. Thomas, 
R. Venugopalan, W. Weise, U. Wiedner, N. Xu, and B. Zou.

The local organizers are \\
A. Dote, Y. Goto, M. Harada, A. Hosaka, K. Itakura, 
H.~Kamano, S.~Kumano (co-chair), A. Monnai, O. Morimatsu, S. N. Nakamura, 
M. Naruki, H. Noumi, H. Ohnishi, K. Ozawa, H. Sako,
F. Sakuma, S. Sawada (co-chair), H. Takahashi, T. Takahashi,
K. Tanaka, and K.~Tanida.

The conference is financially supported by 
APCTP, J-PARC/KEK, JSPS Grant on Innovative Area, 
RCNP at Osaka University, 
Tsukuba Tourism and Convention Association/Tsukuba City.
We also have supports by JAEA, RIKEN and SOKENDAI.

\vfill\eject
\section{Selected topics in the conference}

All the talks cannot be summarized in this article, 
so that selected topics are explained briefly in the following
from discussions at the plenary sessions \cite{qnp2018}.

\subsection{Gravitational waves and nuclear physics}

Neutron-star studies have a long history since the theoretical 
prediction in 1934 \cite{neutron-star-1934}; 
however, there are significant progress
recently in this field motivated by the experimental observations 
on the neutron-star masses and gravitational waves, 
together with theoretical developments on baryon-baryon interactions 
and dense matters \cite{neutron-star-eos}.
The experimental discovery of the gravitational waves 
in 2016 and subsequent developments of this 
gravitational-wave astrophysics have strong impact on nuclear physics.
In particular, there was a first observation (GW170817) on
gravitational waves and electromagnetic radiation in 2017
coming from the merger of two neutron stars \cite{grav-wave}.

Elements beyond the most stable iron are considered to be produced 
mainly by s (slow) and r (rapid) processes.
The r-process elements used to be considered as produced in supernovae,
which were not enough to explain all the existing elements.
However, the neutron-star merger discovery 
paved the way for understanding the generation of the heavy elements.
The r-processes can be inferred by observing the kilonovae, namely
electromagnetic radiations by decays of heavy elements produced 
in the r-processes.
The observations of neutron-star mergers seem to be consistent 
with theoretical kilonovae predictions.
It suggests that neutron-star mergers should be one of major
sources in creating heavy elements by the r-processes.

The neutron-star mergers also provide information on the equation
of state (EOS) \cite{neutron-star-eos}, which is the relation 
between the pressure and density,
for neutron stars by tidal deformability.
Various EOSs have been proposed theoretically
in nuclear physics and they were tested by observing neutron-star
masses and radii. The neutron-star mergers provide 
a constraint on the EOS because the neutron stars are distorted
as they become closer, namely the merging process depends on 
the neutron-star radii and how easy they are stretched.
This EOS affects gravitational-wave amplitude and frequency,
so that it is constrained by the gravitational-wave observations.
So far, the observations are consistent with theoretical models.
Asides from the gravitational waves, we know that
the EOS is significantly softened if hyperons exist
in the neutron stars, so that hyperon-nucleon interactions should be 
investigated experimentally as mentioned in the following 
Sec.\ref{h-int-strange},  such as at 
Japan Proton Accelerator Research Complex (J-PARC),
for providing precise information on building
a realistic theoretical neutron-star model.


\subsection{Hadron interactions and strangeness nuclear physics}
\label{h-int-strange}

The realm of nuclear physics can be expanded by extending 
the flavor degrees of freedom to strange and charm. 
Strangeness nuclear physics is investigated by 
hypernuclei and kaonic nuclei.
The strange-quark mass is of the order of the QCD scale parameter 
$\Lambda_{QCD}$, which indicates that 
the strangeness is appropriate for probing QCD dynamics.
Hyperons ($Y$) could exist in neutron stars, and information
on $YN$ and $YY$ interactions is valuable for the neutron star EOS.
In addition, there is no Pauli blocking for a hyperon  
in a nucleus, so that they are good probes of deep regions of nuclei. 
We started investigating dense many-body systems
without a restriction of the exclusion principle.

One of the major projects at J-PARC
is on strangeness nuclear physics 
by using secondary kaon beams \cite{strange,csb,xi-A,KNN}.
I explain some of major J-PARC results so far.
The first achievement is the discovery of the charge-symmetry
breaking (CSB) in hypernuclei \cite{csb}. The charge symmetry has been taken
as granted in discussing gross nuclear phenomena.
In ordinary nuclei without strangeness, the CSB was observed,
for example, as the binding-energy difference between $^3$H 
and $^3$He. At J-PARC, the energy spacing between 
the spin-doublet states of $^4 _\Lambda$He($1^+$, $0^+$) was 
determined to be $1406\pm 2\pm 2$ keV, which is a clear
indication of the CSB in comparison with 
$^4 _\Lambda$H($1^+$, $0^+$). 
This result sheds light on a new aspect of 
hyperon interactions with $\Sigma \Lambda$ mixture
as the origin of the CSB phenomena in hypernuclei.
Second, from the measurement of the binding energy
($B_{\Xi^-}=4.38 \pm 0.25$ MeV) in the $\Xi^-$-$^{14}$N system \cite{xi-A},
it became clear that $\Xi N$ interaction is attractive. 
It opens a new filed of hypernuclear physics with strangeness $-2$
beyond strangeness $-1$.
Hypernuclei have 
interesting characteristics in experimental observables.
For example, shell structure of each $\Lambda$ state
($s_\Lambda$, $p_\Lambda$, and so on) in nuclei should be
clearly seen at J-PARC by ($\pi$,$K^+$) reactions, much clearly
than observed in a KEK-PS experiment.
Third, there are experimental progress on new hadronic systems
with an antikaon such as $K^- pp$ system \cite{KNN}.
Since an antikaon and a nucleon could form the bound state as
$\Lambda (1405)$, kaonic nuclei could exist.
At this stage, there are large variations in
theoretical and experimental binding energies and decay widths.
However, we expect that future experimental measurements
and theoretical efforts will clarify this issue.
J-PARC strangeness nuclear physics also has
a strong impact on neutron-star EOS.
In the near future, the high-momentum beamline will be ready,
so that the J-PARC physics scope will be extended to different 
nuclear-physics topics.
There is an approved experiment on charmed baryons, so that
charmed-nucleus projects will be possible in future at J-PARC.


Next, short-range nucleon-nucleon correlations have been investigated
in nuclei at 
the Relativistic Heavy Ion Collider (RHIC) and 
Thomas Jefferson National Accelerator Facility (JLab) 
by the reactions $(p,2pN)$ and $(e,e' pN)$, respectively \cite{NN-correlation}. 
The two nucleons are emitted on opposite sides in the final state, so that
these nucleons are considered to be close together in
the initial state, and the process is sensitive to
short-range $NN$ interactions.
It used to be mysterious to find much stronger correlation for $pn$
than $pp$ and $nn$ ones in 2006 \cite{NN-correlation-2006}; however, 
later studies found that it originates 
from the isospin-dependent tensor force. Such isospin dependence affects 
neutron-star studies because protons could exist in the stars 
with a certain fraction \cite{NN-correlation-neutron-stars}.
The short-range correlations are also investigated in the inclusive
electron scattering $A(e,e')X$ at very large $x$ ($>1$), where $x$
is the Bjorken variable.
Especially, the nuclear modification ($\sigma_{^{56}\text{Fe}}/\sigma_{\text{D}}$,
where $D$ is deuteron) plateau was observed at $x=1.4 - 1.9$ 
for the iron to indicate the two-nucleon correlations \cite{NN-correlation}. 
In future, we will step into 
the three-nucleon correlation studies above $x=2$.
It is also possible to investigate such 
three-nucleon correlations at hadron facilities.
The two- and three-nucleon correlation studies at high momenta beyond
the Fermi momentum are valuable for studying dense matters
such as the neutron stars. 


\subsection{Lattice QCD}

Lattice QCD studies became increasingly important 
in nuclear physics with the progress of supercomputer power 
and theoretical developments since it is the only possible 
way to calculate nonperturbative quantities directly from QCD, 
although actual applications are still limited to 
the zero baryon density phenomena.
The lattice studies are in a good situation in hadron spectroscopy,
baryon-baryon interactions, and quark-hadron matters at
the zero density \cite{lattice-QCD}.

Recently, we obtained more measurements on heavy-quark hadrons,
and the lattice QCD is successful in explaining their spectra
below hadronic thresholds. Above the thresholds, simulations
of scattering are necessary. Recently, significant efforts have been made
in lattice QCD to understand the resonances of shallow bound states
and also states, which decay strongly, above the thresholds.
For nuclear physics, the baryon-baryon interaction studies
are valuable. So far, the lattice studies have been limited
to confirming existing phenomenological studies on nucleon-nucleon
interactions determined by scattering experiments and 
deuteron properties. However, the lattice simulations started
to produce predictions recently on phenomena, for which
experimental information does not exist. 
In this sense, lattice QCD is becoming a new stage
where its results could be ahead of phenomenological theory
models and even beyond experimental works, such as
predictions on the bound states of $p \Omega^-$ and 
$\Omega^- \Omega^-$ systems.
Since we know that QCD is the fundamental theory for strong interactions,
even experimental measurements may not be necessary if QCD {\it were} 
precisely solved by the lattice simulation including finite densities.

The fast-developing area of lattice QCD in hadron physics
is on an application to parton distribution functions (PDFs) 
\cite{lattice-QCD}.
Since they are defined by lightcone-separated field correlators,
it was not possible to calculate the Bjorken-$x$ dependent PDFs.
However, we may study quasi-PDFs with an equal-time separation
so that it becomes possible to calculate the quasi-PDFs by lattice QCD.
The quasi-PDFs agree with the PDFs in the infinite momentum limit 
$p_z \to \infty$, which is very challenging numerically.
There are already numerical results on some PDFs.
This project is valuable particularly for the PDFs,
where there is little experimental information. 


\subsection{Hadron spectroscopy}

Hadron spectroscopy has been investigated since 1960's.
However, the last decade is an especially appropriate time 
to extend basic quark-model descriptions
in the sense that there are many reports 
on exotic-hadron candidates \cite{spectroscopy}. 
The name ``exotic hadron" has been used for a hadron with
a different internal configuration from $q\bar q$ and $qqq$ 
in the basic quark model.
Experimental efforts have been made to find deviations from
theoretical calculations of phenomenological 
models which assume the $q\bar q$ and $qqq$ configurations.
On the other hand, because the quark number is not a conserved 
quantity in QCD, we should think what the exotic hadrons really mean.
For example, since valence-quark numbers are related to
the conserved quantities such as charges and baryon numbers,
and since constituent quark numbers are counted in hard exclusive 
reactions according to perturbative QCD, there may be a way to use
high-energy hadron reactions to find internal structure
of exotic hadron candidates \cite{counting}.
The 3D tomography technique in Sec.\,\ref{nucleon} could be also
used in future for clarifying the internal configuration.

For understanding confinement in QCD, excited states of the nucleon (N$^*$) 
have been investigated. The N$^*$ program at electron
accelerator facilities, such as JLab, has two major projects.
One is to measure the N$^*$ spectrum and decay widths systematically,
and the other is to study transition form factors from the nucleon 
to the excited states, for understanding low-energy QCD
and confinement.
There are significant improvements for N$^*$ states above the mass of 1700 MeV
from the PDG (Particle Data Group)-2012 version by confirming them
as 4-star (certain existence) states.
In the last several years, there were discovery reports especially from 
BaBar, BES, Belle, and LHCb for exotic-hadron candidates
in heavy hadron systems called $X$, $Y$, and $Z$ states
with charm and bottom.
Such exotic candidates are, for example, 
$X (3872)$, $Y(4260, 4360, 4660)$, $Z_c (3900, 4020, 4200, 4430)$, 
$Z_b (10610, 10650)$, and $P_c (4380, 4450)$ \cite{spectroscopy}. 
These states could be multiquark states, hadron molecules,
and/or mixture of these states.
Kinematical cusp effects should be also considered
in discussing the exotic-hadron candidates.
There are also recent progress on heavy-quark hadrons, namely
excited charmed baryons of $\Omega_c$,
doubly-charmed baryons $\Xi_{cc}$, 
other doubly-heavy-quark baryons,
and bottom baryons $\Xi_b$ and $\Sigma_b$.
There are theoretical predictions on 
doubly-heavy tetraquark hadrons.
On the other hand, diquark degrees of freedom could become 
important in excited baryon spectra, and it could be seen
in baryons with a heavy quark like charm or bottom quark \cite{e50}.


\subsection{Nucleon structure}
\label{nucleon}

The project on unpolarized structure functions of the nucleon is becoming
one of precision-physics fields because of accurate determination
of the PDFs and theoretical efforts on higher-order perturbative-QCD
corrections \cite{sf-summary}.
It is the basics for finding a new hadron-physics 
phenomenon and a signature beyond the standard model 
in any high-energy hadron reactions, especially at 
the Large Hadron Collider (LHC).
On the other hand, the issue on the origin of nucleon spin 
has not been clarified yet \cite{sf-summary,nucleon-sfs,spin-decomposition}.
Theoretical efforts had been done on how to decompose the nucleon
spin into quark and gluon spin and orbital-angular-momentum (OAM)
contributions in a color-gauge invariant way. This theoretical
issue was settled down recently \cite{spin-decomposition}. 
The remaining work is to determine gluon-spin and partonic OAM effects 
in experiments. Especially, in order to determine the OAM contributions, 
we need to investigate three-dimensional (3D) structure functions
including transverse structure.
This field is called hadron tomography,
whereas one-dimensional structure functions, expressed
by the longitudinal-momentum fraction $x$, have been mainly
investigated.

There are three types in the 3D structure functions:
generalized parton distributions (GPDs),
generalized distribution amplitudes (GDAs), and
transverse-momentum-dependent parton distributions (TMDs)
\cite{sf-summary,nucleon-sfs}.
The second moments of the GPDs are related to the total
angular momentum carried by partons, so that they are
key quantities to solve the nucleon spin puzzle.
The GPDs contain the PDFs and transverse form factors.
There are experimental projects by JLab 
and CERN-COMPASS and in future by EIC.
The GDAs are $s$-$t$ crossed quantities of the GPDs, and
they are studied by KEKB and in future by 
GSI-FAIR (Gesellschaft f\"ur Schwerionenforschung -Facility for 
Antiproton and Ion Research).
The GPDs and GDAs are related to hadronic matrix elements
of energy-momentum tensor of quarks and gluons,
and they are expressed by gravitational form factors.
Therefore, this field is related to the origin of hadron
masses and also pressure/shear-force in hadrons
in terms of quark and gluon degrees of freedom.

The TMDs are or will be investigated by 
COMPASS, Fermilab, GSI-FAIR, LHC, JLab, and RHIC.
There is an interesting development recently on explicit
phenomenon of color flow in the TMD studies.
The color flow is considered in structure functions
to satisfy the color gauge invariance and it is called
gauge link in lattice QCD. In the TMDs, the color flow
should exist in transverse directions in addition
to the longitudinal one, so that the TMDs could change
a sign depending on processes. Recently, this color-flow
effect was found by observing the TMDs
in COMPASS and RHIC experiments. 
It could open a new field on explicit
color phenomena in hadron physics.
The field is also connected to color versions of 
quantum entanglement and the Aharonov-Bohm effect
\cite{sf-summary,ab-qcd}.


\subsection{Hot and dense quark-hadron matters}

Soon after the big bang, the universe was filled with hot and 
dense particles, which were dominated by quarks and gluons, and then
they formed hadrons and nuclei as illustrated in Fig.\,\ref{fig:qcd-phase}.
High-energy heavy-ion collision experiments
are intended to investigate QCD at finite density and temperature 
by laboratory experiments \cite{heavy-ion-medium}.
High-energy experiments are going on at LHC and RHIC.
The initial state of high-energy heavy-ion collisions is described by 
dense gluon systems called color glass condensate.
Then, these two Lorentz-contracted gluon sheets pass through each other
to produce longitudinal electric and magnetic fields, 
and this matter at the pre-equilibrium stage is called glasma.
Subsequently, the system reaches a thermodynamic equilibrium
as quark-gluon plasma, which is described by hydrodynamics. 
Finally, hadronization occurs and then hadrons are observed
by experiments. Theorists have been working
a comprehensive framework based on transport theory for
describing the reaction from the initial to the final state.

There was an important discovery at RHIC to find a nearly-perfect 
fluid for the quark-gluon plasma by observing the elliptic flow
in comparison to hydrodynamical calculations
\cite{heavy-ion-medium}.
However, recent experimental measurements indicate that the elliptic
flow exists even for proton-proton (pp) and proton-nucleus (pA) collisions.
Therefore, we need to understand the phenomena first for reactions of
small systems such as pp and pA, and then the AA collisions may be 
reinvestigated.
Next, since the largest magnetic field in nature is created
in heavy-ion reactions, it is interesting to investigate topics
associated with it. Quarks have spin, and their chiral magnetic effects
are now under serious investigations. The effects should be observed as
azimuthal correlations of produced particles in the final state.
The topic of the chiral magnetic effects is fast developing
as an interdisciplinary field such as with 
neutrino spin in neutron stars for supernova 
and with Weyl semimetals in condensed-matter physics.


Hadron masses in nuclear medium have been investigated to
find the origin of hadron masses \cite{heavy-ion-medium}.
The up- and down-quark masses are much smaller than hadron masses, 
so that we need to find how the major part 
of the hadron masses is generated. 
The traditional idea is due to chiral symmetry breaking.
An order parameter of the symmetry breading is the quark condensate 
$\langle q\bar q \rangle$. Since it is not an observable, meson-mass shifts,
which are connected to the condensate, are experimentally
investigated in nuclear medium.
Experimental studies have been done at
CERN-SPS, COSY, ELSA, GSI-HADES, JLab, KEK-PS, MAMI, RHIC, and SPring-8.
In order to find the chiral-symmetry breaking from experimental 
measurements, we need to have transport-model calculations
to compare with experimental observables.
We also would like to have consistent measurements on mass shifts and
widths among experimental groups. This topic should be investigated by
the future experimental projects at 
CERN-SPS, GSI-FAIR, HIAF (High-Intensity Heavy Ion Accelerator Facility), 
J-PARC, and NICA (Nuclotron-based Ion Collider fAcility).


\subsection{New experimental facilities}

I briefly explain some of recent and future facilities on 
nuclear physics related to QNP2018.
It should be noted that the following is not a complete list.
For example, COMPASS, HIAF, NICA, and some other facilities are not mentioned.

\vspace{-0.20cm}
\subsubsection{J-PARC}

The J-PARC is a high-intensity accelerator
of the 1 MW range at relatively high energies 3-30 GeV
\cite{j-parc}.
Material and life-science experiments are done by using 
neutrons and muons produced by the 3-GeV proton beam.
Nuclear and particle physics are investigated
with secondary beams as well as the primary-proton beam.
The relevant nuclear-physics experiments for the QNP2018
are done in the hadron experimental facility, where
secondary beams such as pions, kaons, and antiprotons
as well as the primary 30-GeV proton beam are available.
The hadron experiments are intended to study
new forms of hadronic and nuclear systems by extending
flavor degrees of freedom to strange.
The high-momentum beamline will be ready soon,
so the J-PARC project will be extended to other hadron topics
such as hadron properties in nuclear medium, charmed baryons,
and nucleon structure. There is also a proposal
on heavy-ion acceleration to study quark-hadron matters.
The first priority of future hadron physics projects
is to extend the current hall for 105 m in length 
for full operation of hadron and nuclear physics projects.

\vspace{-0.20cm}
\subsubsection{JLab}

In 2017, the JLab completed the upgrade of its Continuous Electron 
Beam Accelerator Facility to achieve the beam energy of 12 GeV
\cite{jlab}, 
the construction of the new end station Hall D, and new detector equipments.
The project includes a wide range of topics, including
the unpolarized and polarized 3D quark distributions in the nucleons 
and nuclei, as well as the longitudinal PDFs, at medium and large $x$
by deeply exclusive and semi-inclusive electron-scattering processes.
The GPD studies will establish 3D picture of the nucleon including
the transverse coordinates, and they are connected to the gravitational
form factors. The TMD studies could allow us to step into 
the color-flow physics.
The color confinement will be investigated by hadron spectroscopy
and electromagnetic form factors of nucleons and nucleon resonances.
Physics beyond the standard model will be investigated by parity-violating 
electron scattering. The JLab will play a leading role 
in hadron physics at medium and high energies
in the next decade.

\vspace{-0.20cm}
\subsubsection{Super-KEKB}

The Super-KEKB factory is an electron-positron collider 
with the center-of-mass energy of 10.6 GeV and the peak luminosity 
$8 \times 10^{35} \text{cm}^{-2} \text{s}^{-1}$, which is 40 times larger 
than the KEKB collider \cite{kekb}. It was completed in 2018.
Its major motivation is to study CP violation and 
flavor physics to find new physics beyond the standard model.
However, many new announcements of KEKB have been on discoveries 
of new hadrons.
In fact, the $X(3872)$ article has the highest citations of 1571
in all the papers published by the Belle collaboration of KEKB, 
so that it is the most famous discovery of KEKB so far.
The KEKB has significant contributions
not only in the discoveries of exotic-hadron candidates
named $X$, $Y$, and $Z$ but also in precision determination 
of fragmentation functions. 
The Super-KEKB facility will produce measurements 
with much better precisions, and we expect to have more discoveries 
and precision measurements on exotic hadrons and fragmentation functions.
In addition, the GDAs are $s$-$t$ crossed quantities of the GPDs, and they
are investigated by the two-photon processes at KEKB. Such studies lead
to the understanding of 3D structure of hadrons and their gravitational
form factors \cite{hadron-mass-1}.

\vspace{-0.20cm}
\subsubsection{GSI-FAIR}

The GSI-FAIR is a multipurpose 
facility with four experimental collaborations:
APPA (Atomic, Plasma Physics and Applications), 
CBM (Compressed Baryonic Matter),
NUSTAR (Nuclear Structure, Astrophysics and Reaction), 
and PANDA (Antiproton Annihilation at Darmstadt) \cite{gsi-fair}.
The APPA is for fundamental investigations on material sciences
and biophysics including medical applications.
The CBM is for understanding dense and hot nuclear matter
at lower temperature and higher baryon density than RHIC and LHC.
The NUSTAR is for investigating exotic nuclei and heavy elements 
far off stability to understand nucleosynthesis.
The PANDA is for hadron structure and dynamics with antiproton beams.
The FAIR will provide beams of ions and antiprotons. 
The existing facility including SIS18 has been upgraded 
to serve as injector for FAIR, and it is considered as
the FAIR phase-0 program.
The main component of FAIR is the SIS100 ring accelerator, and it will
accelerate protons up to 29 GeV (4 GeV at SIS18) 
and heavy ions up to 11 GeV/u (1 GeV/u).
The facility is now under construction and it is expected 
to be completed in 2025.

\vspace{-0.20cm}
\subsubsection{EIC}

An Electron-Ion Collider (EIC) is a major future nuclear-physics 
program in US. There are other EIC projects at CERN and in China, but 
I focus on the US facility in the following. 
There are two site possibilities at BNL and JLab.
The project is in progress for completing it in the middle of 2020's.
The c.m. energies span 22$-$63 GeV (15$-$40 GeV) or 45$-$141 GeV (32$-$90 GeV)
for polarized $ep$ ($eA$) collisions \cite{eic}. It will extend 
the current kinematical range to smaller $x$, as small as $10^{-4}$, 
with higher $Q^2$.
The physics motivation is to understand how quarks and gluons make up 
nearly all of the visible matter in the universe. 
We expect that it will be realized by investigating
3D tomography of nucleons and nuclei, solving the proton spin puzzle, 
finding color-glass condensate, and studies of quark and gluon confinement.
The EIC was included in the 2015 Nuclear Science Advisory Committee (NSAC) 
Long Range Plan. 
In 2018, a report by the National Academies of Sciences, Engineering, 
and Medicine positively endorsed the EIC proposal.
We expect that it will be the leading facility from the middle of 2020's
to clarify basic issues of hadron physics.

\section{Conclusion}

Nuclear physics is the field of investigating matter generation of universe
and properties of quark-hadron many-body systems as ultimate materials.
We have excellent opportunities now to enhance our activities,
not only because of major experimental facilities: 
BES, COMPASS, EIC, Fermilab, GSI-FAIR, HIAF, 
KEKB, JLab, J-PARC, LHC, NICA, RHIC, 
and so on,
but also relations with other developing fields, such as 
gravitational waves, quantum computation/entanglement, 
high-energy cosmic rays, neutrino physics, and others.
The hadron-nuclear physics needs to be developed together 
with neighboring fields, especially condensed-matter physics, 
particle physics, and astrophysics.
There are 89 young students and postdocs 
among the total participant number of 216 at this QNP2018
conference, which suggests a bright future of our field.

\section*{Acknowledgments}

The author thanks K. Kyutoku  and A. Monnai for useful comments
in writing this article.

\vfill\eject


\end{document}